\documentclass[preprint,prd]{revtex4}

\usepackage{amsbsy}
\usepackage{amssymb}
\usepackage{amsmath}
\usepackage{graphicx}

\def\x{{\mathrm{x}}}
\def\y{{\mathrm{y}}}

\def\n{{\rm n}}
\def\p{{\rm p}}
\def\e{{\rm e}}

\def\bBp{\bar{\mathcal{B}}'}
\def\bB{\bar{\mathcal{B}}}
\def\bep{\bar{\varepsilon}}

\def\be{\begin{equation}}
\def\ee{\end{equation}}
\def\bea{\begin{eqnarray}}
\def\eea{\end{eqnarray}}

\def\bear{\begin{eqnarray}}
\def\eear{\end{eqnarray}}

\begin{document}

\title{Superfluid instability of r-modes in ``differentially rotating'' neutron stars}

\author{N. Andersson$^1$, K. Glampedakis$^{2,3}$ \& M. Hogg$^1$}

\affiliation{
$^1$School of Mathematics, University of Southampton,
Southampton SO17 1BJ, United Kingdom\\
$^2$Departamento de Fisica, Universidad de Murcia, E-30100 Murcia, Spain\\
$^3$Theoretical Astrophysics, University of T\"ubingen, Auf der Morgenstelle 10, T\"ubingen, D-72076, Germany}

\begin{abstract}
Superfluid hydrodynamics affects the spin-evolution of mature neutron stars, and may be key to explaining timing irregularities such as 
pulsar glitches. However, most models for this phenomenon exclude the global instability required to trigger the event.
In this paper we discuss a mechanism that may fill this gap.
We establish that small scale inertial r-modes become unstable in a superfluid neutron star that exhibits
a rotational lag, expected to build up due to vortex pinning as the star spins down. Somewhat counterintuitively, this instability arises due to the (under normal circumstances dissipative) vortex-mediated mutual friction. 
We explore the nature of the superfluid instability for a simple incompressible model, 
allowing for entrainment coupling between the two fluid components. Our results recover a previously discussed dynamical instability in systems where the two components are strongly coupled. In addition, we demonstrate for the first time that the system is secularly unstable (with a growth time that scales with the mutual friction) throughout much of parameter space. Interestingly, large scale r-modes are also affected by this new aspect of the instability. We analyse the damping effect of shear viscosity, which should be particularly efficient 
at small scales, arguing that it will not be sufficient to completely suppress the instability  in astrophysical systems. 
\end{abstract}

\maketitle

\section{Introduction}

Since the first observation of a glitch in the spindown of the Vela Pulsar in 1969 \cite{vela1,vela2}, there have been a large number of similar events observed in other pulsars \cite{glitch,espinoza}. Several competing,  in some cases complimentary, mechanisms have been suggested as explanation for these occurrences \cite{sidery,pizzochero,warsaw1, warsaw2,haskell}. The two most widely classes models relate either to changes in the star's elastic crust or the dynamics of  the superfluid neutrons that are present both in the core and  the inner crust. The first set of models involve fractures in the crust, leading to changes in the moment of inertia of the star \cite{crust}. The second set involves the unpinning of vortices associated with a superfluid component from the star's inner crust \cite{pinned}. The unpinning event is attributed to the build-up of ``differential 
rotation'' between the two components (elastic crust and interpenetrating superfluid). It is the second of these two mechanisms with which we concern ourselves in this paper.

We consider a promising mechanism for triggering the observed events; a superfluid instability present in systems with ``differential rotation'', like the lag that plays a central role in all vortex-based glitch models. The basic mechanism has already been discussed in \cite{GAprl}, where it was argued that 
the instability may trigger large-scale vortex unpinning leading to the observed events. The first part of the argument was a demonstration that  unstable r-modes could be generated by the difference in rotation of the charged particles (protons and electrons) and the superfluid neutrons in the star. In particular, it was shown in \cite{GAprl} that such unstable modes exist for a strongly coupled system. The second part of the argument consisted of a demonstration that viscosity would not suppress this instability completely, but would allow growing modes to exist for a range of wavelengths once a critical rotational lag was reached. The third part of the argument placed the mechanism in an astrophysical context by comparing the predictions of the model to 
observational data.

In this paper we consider this new r-mode instability in more detail. We extend the analysis of \cite{GAprl} beyond the strong-drag limit, and show that  dynamically unstable modes (with a growth time similar to the star's rotation rate) exist for a  wide range of parameter values. In addition, we discover that the r-modes also suffer a secular instability (with growth time proportional to the mutual friction) throughout much of parameter space. This aspect is new. It is particularly interesting as it affects also the  global scale r-modes, while the dynamical instability is restricted to small scale modes. As in the previous work, we limit ourselves to discussion of the non-relativistic case. Again following the approach of \cite{GAprl}, we introduce shear viscosity and demonstrate that the instability is not completely suppressed by the inclusion of the associated damping.

The main purpose of this investigation is to establish the robustness of the superfluid instability, highlight the key ingredients that lead to its presence and set the stage for more detailed studies of this mechanism. The instability that we consider may be generic (belonging to the class of two-stream
instabilities that are known to operate in a wide variety of physical systems \cite{twostream}), but at this point it is not clear how it is affected by  other aspects of 
neutron star physics. In particular, future work needs to extend the analysis to account for the crust elasticity and the presence of the star's magnetic field. 

\section{Setting the stage}

We consider the conditions that are expected to prevail in the outer core of a mature neutron star, where a neutron superfluid is thought to coexist with a 
proton superconductor. A key qualitative aspect of this system is that it allows the two ``fluids'' to flow ``independently", leading to dynamics 
that is well represented by a two-fluid model. In order to keep the analysis manageable, we assume that both fluid components are charge neutral.
This allows us to neglect (in this first proof-of-principle analysis) the star's magnetic field. This is obviously a severe simplification, but to analyse the 
full problem would be difficult at this point. The key aspect that we add to previous studies in this problem area is the relative rotation between 
the two components,  expected to build up as the observed component spins down due braking associated with  the exterior magnetic field. We are interested in the global oscillations of 
a star that exhibits this kind of ``differential'' rotation. Despite this being a key aspect of the generally accepted ``explanation'' for observed pulsar glitches, there have not yet been any detailed studies of the dynamics of such a system. In this sense, the present work provides an important step towards realism. 

The main question that motivates us stems from the discovery that two-stream instabilities are generic in the two-fluid model \cite{2stin1}. Given this, it is relevant to ask whether the rotational lag that builds up in a mature neutron star may lead to such instabilities and, if so, what the 
observational consequences may be. Conversely, given that we do not yet know what the mechanism that triggers the observed glitches may be, 
but that it ought to be some kind of ``global'' instability, it is interesting to ask whether the two-stream instability may be relevant in this context. 

We consider the linear perturbations of a system that exhibits a rotational lag. Because of this set-up, it is natural to 
work in the inertial frame (rather than choosing one of the rotating frames). It is also natural, given the complex nature of the perturbed velocity fields etcetera to carry out the analysis in a coordinate basis. 
This means that vectors, like the velocity, are expressed in terms of their components, $v^i$ (say), and  a distinction is made between 
co- and contravariant objects, with the former following from the latter as $v_i=g_{ij}v^j$ where $g_{ij}$ is the flat three-dimensional
metric. This description of the problem
is also advantageous since it involves the use of the covariant derivative $\nabla_i$ associated with the given metric, which automatically encodes the scale factors associated with the curvilinear coordinates ($r,\theta,\varphi$) that are appropriate for the problem.

We consider a system with two interpenetrating fluids, labeled $\n$ and $\p$ (from now on),  which are assumed to rotate rigidly in such a way that
\be
v_\x^i = \Omega_\x \hat{e}_\varphi^i \qquad \ , \qquad \x = \n,\p
\ee
where $\hat{e}_\varphi^i$ represents the azimuthal basis vector (and $\hat{e}_\theta^i$ later the polar one). The angular velocities, $\Omega_\x$, are assumed to be constant, but we  allow for differential rotation, i.e. $\Omega_\n^i \neq \Omega_\p^i$. To keep the analysis tractable, we assume that the
two rotation axes coincide, and use
\be
\Omega_\p^i = \Omega^i \ , \qquad \mbox{and} \qquad \Omega_\n^i = (1+\Delta)\Omega^i
\ee
Making contact with pulsar observations, $\Omega^i$ would be the observed (angular) rotation frequency (i.e. that of the crust) while $\Omega_\n^i$ represents the rotation of the unseen (interior) superfluid component. As the superfluid is expected to lag behind as the crust spins down, 
one would typically expect $\Delta$ to be small (of the order of $10^{-4}$ at the time of a glitch \cite{glitch}) and positive. 

Key to understanding the dynamics of glitching pulsars is the appreciation that the neutrons may form Cooper pairs, and hence act as a superfluid. 
The upshot of this is that bulk rotation can only be achieved by the formation of an array of quantized vortices. As discussed in, for example, 
\cite{supercon} the quantization condition is imposed on the momentum that is conjugate to the velocity $v_\n^i$. This momentum 
is given by
\be
p^\n_i = m_\p \left( v_i^\n + \varepsilon_\n w_i^{\p\n}\right) 
\ee
where $m_\p$ is the nucleon mass (we ignore the small difference between the bare neutron and proton masses),  $\varepsilon_\n$ represents the entrainment between neutrons and protons (more of which later) and the relative velocity is
\be
w^i_{\y\x} = v^i_\y-v^i_\x \ , \qquad \y\neq \x
\ee 
Given this, and the relevant quantization condition (see \cite{supercon} for discussion),
the local vortex density (per unit area), $n_v$, follows from 
\be
n_v \kappa^i = \epsilon^{ijk}\nabla_j [v^\n_k + \varepsilon_\n w^{\p\n}_k)] =
2[\Omega_\n^i + \varepsilon_\n(\Omega_\p^i-\Omega_\n^i)]= 2[1+(1-\varepsilon_\n)\Delta]\Omega^i
\label{quant} \ee
where  $\kappa^i$ is the vector aligned with the vortex array with magnitude $\kappa=h/2m_\p$ ($h$ is Planck's constant).
Here we have introduced yet another simplifying assumption; we have taken the fluids to be incompressible which  implies that   the entrainment 
coefficient may (at least in the small $\Delta$ case that we are focussing on) be taken to be constant. It is also worth noting that
$\rho_\n \varepsilon_\n = \rho_\p\varepsilon_\p$  where $\rho_\x=m_\p n_\x$ are the respective mass densities.

Let us now consider the linear perturbations (represented by $\delta$) of this kind of configuration. Assuming that the flow is incompressible also at this level 
(incidentally not a bad approximation for the inertial flows that we will consider) we have, first of all, 
\be
\nabla_i \delta v_\x^i = 0
\ee
Meanwhile, the perturbed momenta 
\be
\delta p^\x_i = \delta v^\x_i + \varepsilon_\x \delta w^{\y\x}_i
\ee
 are governed by the Euler equations;
\be
\mathcal{E}^\x_i = i\omega \delta p^\x_i + \delta v_\x^j \nabla_j p^\x_i + v_\x^j \nabla_j \delta  p^\x_i
+ \varepsilon_\x (\delta w^{\y\x}_j \nabla_i v_\x^j +  w^{\y\x}_j \nabla_i \delta v_\x^j) +
\nabla_i \delta \Psi_\x
 = \delta f^\x_i
\label{eulers}\ee
where
\be
\Psi_\x = \Phi + \tilde{\mu}_\x
\ee
combines the gravitational potential $\Phi$ and the chemical potential $\tilde\mu_\x=\mu_\x/m_\p$ for each fluid. We have assumed that the time-dependence is harmonic, $\propto \exp(i\omega t)$, as we are interested in the oscillation modes of the system. Later, we will make extensive use of the equations that govern the perturbed vorticity. Specifically, 
we will work with the  quantity
\be
\mathcal{W}_\x^i = \epsilon^{ijk}\nabla_j \mathcal{E}^\x_k
\ee

The right-hand side of \eqref{eulers} can be used to account for any ``external'' forces acting on the fluids. It can also be used to 
incorporate  interactions between them. The main such interaction force is the mutual friction that arises due to the presence of the 
quantized vortices \cite{alpar,mendell}. The unperturbed force takes the form \cite{trev}
\be
f_i^\x = { \rho_\n \over \rho_\x} \mathcal{B}' n_v \epsilon_{ijk} \kappa^j w_{\x\y}^k
+  { \rho_\n \over \rho_\x} \mathcal{B} n_v \epsilon_{ijk} \hat{\kappa}^j \epsilon^{klm}\kappa_l w_m^{\x\y}
\label{fmf}\ee
where $\mathcal{B}$ and $\mathcal{B}'$ are coefficients to be determined from microphysics. In the standard scenario these coefficients can be expressed in terms of a single resistivity $\mathcal R$, associated with scattering off of the vortices, such that \cite{trev}
\be
\mathcal{B} = {\mathcal{R} \over 1 + \mathcal{R}^2}
\ee
and
\be
\mathcal{B}' =  {\mathcal{R}^2 \over 1 + \mathcal{R}^2}
\ee
The weak- and strong coupling limits discussed later correspond to, respectively, $\mathcal R \to 0$ and $\mathcal R \to \infty$. It is worth noting that the system can build up a differential rotation lag only as long as some additional force (like vortex pinning) prevents 
the mutual friction from acting. The  presence of such a force is implicitly assumed in the following discussion. 

Perturbing (\ref{fmf}), we get
\begin{multline}
\delta f_i^\x =  { \rho_\n \over \rho_\x} \mathcal{B}' \left[ \delta (n_v \kappa^j) \epsilon_{ijk} w_{\x\y}^k +
n_v \kappa^j \epsilon_{ijk} \delta w_{\x\y}^k \right] \\
+ { \rho_\n \over \rho_\x} \mathcal{B} \epsilon_{ijk}\epsilon^{klm} \left[
\delta (n_v \kappa_l) \hat{\kappa}^j w_m^{\x\y} +
n_v \kappa_l ( w_m^{\x\y} \delta \hat{\kappa}^j +
 \hat{\kappa}^j \delta w_m^{\x\y})
\right]
\end{multline}
To evaluate this we need 
\be
\delta (n_v \kappa^i) = \epsilon^{ijk} \nabla_j \delta p^\n_k
\ee
and
\be
\delta \hat{\kappa}^i = { 1 \over n_v \kappa} \left(\delta^i_j  - \hat{\kappa}^i \hat{\kappa}_j \right) \epsilon^{jlm} \nabla_l \delta p^\n_m
\ee
where $\hat \kappa^i$ is a unit vector aligned with $\kappa^i$,
together with, cf. \eqref{quant},
\be
n_v \kappa = 2 [ 1 + (1 - \varepsilon_\n) \Delta] \Omega
\ee
In the perturbation equations discussed in the next section, we will not work with the momentum equations \eqref{eulers} directly. 
Rather, we use two combinations that represent the total (perturbed) momentum and the difference. It is well established that these 
combinations isolate the two dynamical degrees of freedom in an ``uncoupled'' two-fluid system \cite{anco}. Hence, this decomposition is often used in 
studies of oscillating superfluid neutron stars \cite{2fosc}. This means that we need;
\be
\rho_\n \delta f_i^\n + \rho_\p \delta f_i^\p = 0
\ee
and
\be
 \delta f_i^\p -  \delta f_i^\n = - { 1 \over x_\p}  \delta f_i^\n
\ee
where we have introduced the proton fraction $x_\p=\rho_\p/(\rho_\n +\rho_\p)$.
Only the second degree of freedom is explicitly affected by the mutual friction.

Finally, we  want to account for the presence of shear viscosity (in the superfluid case mainly due to electron-electron scattering). 
In the incompressible case, this means that we add a force to (the right-hand side of) the proton equation of form; 
\be 
\delta{f}_{\rm sv}^i = \nu_{\rm ee} \nabla^2 \delta {v}^i_\p
\ee
where $\nu_{\e\e}$ is the kinematic viscosity coefficient.
We ignore bulk viscosity for two reasons. First of all, glitching pulsars are cold enough that shear viscosity should be the dominant damping mechanism. Secondly, the particular class of fluid motion that we consider is not efficiently damped by bulk viscosity, anyway \cite{rmode}.

\section{The axial perturbation equations}

Despite the  various simplifying assumptions, the general perturbation problem is challenging.  
It is well-known that, even in the slow-rotation limit where the star remains spherical, there exists a class of inertial modes
\cite{lock} which require the coupling of many (spherical harmonics) multipoles for their description. We are, however, not going to attempt 
to solve the general problem. Instead we will ask a very specific question; How does the presence of the rotational lag affect the 
r-modes of the system? This question is relevant for a number of reasons, perhaps the most important being related to the fact that the 
large scale r-modes may be driven unstable by the emission of gravitational radiation \cite{rmode}. From a practical point-of-view, it is natural to focus on the r-modes since they are associated with particularly simple velocity fields. The hope would be that the corresponding problem remains tractable even when we add the 
rotational lag.

The r-modes are a subclass of inertial modes (restored by the Coriolis force) that are purely axial/toroidal to leading order. This means that 
the perturbed velocities take the form
\be
\delta v_\x^i = -{im \over r^2 \sin \theta}  U^l_\x Y_l^m\hat{e}_\theta^i
+{1 \over r^2 \sin\theta}  U^l_\x \partial_\theta Y_l^m \hat{e}_\varphi^i
\ee
where $Y_l^m(\theta,\varphi)$ are the standard spherical harmonics, $U_\x^l(r)$ are the mode amplitudes, and the symmetry of the problem is such that the $m$-multipoles (proportional to $e^{im\varphi}$) decouple. In the single (barotropic) fluid case, r-mode solutions exist for each individual $l=m\ge 1$. The main focus in previous work has been on the quadrupole mode ($l=2$) since it is associated with the fastest growing gravitational-wave instability \cite{rmode}. In this work we will end up studying a  large range of  scales, including small scale modes with $l$ the order of 100. In a neutron star, such perturbations would be relatively local, corresponding to a typical lengthscale of 100~m. 

As hinted at in the previous section, 
we prefer to work with slightly different perturbation variables. Experience from other problems involving the two-fluid model \cite{2fosc} suggests that 
it is advantageous to work with the total perturbed momentum flux, defined as
\be
\rho U^l = \rho_\n U^l_\n + \rho_\p U^l_\p
\ee
with $\rho=\rho_\n+\rho_\p$. We take the second variable to be given by 
the velocity difference;
\be
u^l = U^l_\p -U^l_\n
\ee
In order to ``simplify'' the final perturbation equations it is useful to express the frequency in the ``rotating frame'', i.e. work with $\sigma$ defined by 
\be
\omega + m\Omega = \sigma  \Omega
\ee
It is also convenient to introduce
\be
L= l(l+1)
\ee
\be
Z = 1 - x_\p - \varepsilon_\p
\ee
and
\be
\bar{Z} = {x_\p Z \over 1-x_\p}
\ee
Finally, we use (as in \cite{2fosc}) the scaled mutual friction coefficients;
\be
\bBp = {\mathcal{B}}'/x_\p \ , \qquad \mbox{ and } \qquad \bB = {\mathcal{B}}/x_\p
\ee

The perturbation equations can now be obtained by inserting the expected velocity field in the equations from the previous section. After some 
manipulations this leads to the following set of the equations to be solved:
From $\rho_\n \mathcal{W}_\n^r + \rho_\p \mathcal{W}_\p^r$ we, first of all, get
\be
\left[ L\sigma - 2m \right] U^l Y_l^m = - m\Delta (L-2)[ U^l- \bar{Z} u^l ] Y_l^m
\ee
(Here, and in the following, summation over $l\ge m$ is implied.)
The advantage of working with this combination, that represents to total vorticity, is that  there are no
mutual friction terms in the equation. Such terms are  associated with the relative flow. 
To see this we consider the combination $ \mathcal{W}_\p^r - \mathcal{W}_\n^r$, leading to
\begin{multline}
\left[ {L\sigma \bar{Z} \over x_\p} - 2m(1-\bBp)- 2 i\bB (L-m^2) \right] u^l Y_l^m = \\
= -  m\Delta \left\{  { [(L-4)x_\p + 2] \bar{Z} \over x_\p} -2(1 - x_\p) \right\} u^l Y_l^m
+ m\Delta \left(L-{ 2 \bar{Z} \over x_\p} \right) U^l Y_l^m \\
+m \Delta L \bBp (\bar{Z}u^l - U^l) Y_l^m - 2m\bBp\Delta (\bar{Z}+1-x_\p)u^l Y_l^m \\
+i \bB \Delta L [ \bar{Z}(r \partial_r u^l - u^l ) - r \partial_r U^l + U^l ] Y_l^m
+2i\bB \Delta (L-m^2) (\bar{Z} +1 - x_\p) u^l Y_l^m \\
+i\bB L \Delta [ r\partial_r U^l - 3U^l - \bar{Z} (r \partial_r u^l - 3 u^l )]\cos^2 \theta Y_l^m \\
+i\bB \Delta [ 2r\partial_r U^l - LU^l - \bar{Z} (2r\partial_r u^l - L u^l) ] \cos \theta \sin \theta \partial_\theta Y_l^m
\end{multline}

We will also use the radial components of the Euler equations. From the combination $\rho_\n \mathcal{E}_\n^r + \rho_\p \mathcal{E}_\p^r$
we find
\begin{multline}
[ x_\p  r\partial_r \delta \Psi_\p^l + (1-x_\p) r\partial_r \delta \Psi_\n^l ] Y_l^m
- 2\Omega U^l \sin \theta \partial_\theta Y_l^m = \\
= - \Delta \Omega (1-x_\p) [ (x_\p -\bar{Z}) r \partial_r u^l + 2 \bar{Z} u^l - 2U^l] \sin \theta \partial_\theta Y_l^m
\end{multline}
while $\mathcal{E}_\p^r - \mathcal{E}_\n^r$ leads to
\begin{multline}
(r\partial_r \delta \Psi_\p^l - r \partial_r \delta \Psi_\n^l) Y_l^m - 2\Omega (1-\bBp) u^l  \sin \theta \partial_\theta Y_l^m 
- 2im \bB \Omega u^l \cos \theta Y_l^m =  \\
= -\Delta \Omega \Bigg\{ \left( 1 - { \bar{Z} \over x_\p} \right) r \partial_r U^l + { 2\bar{Z} \over x_\p} U^l
+ (1-2x_\p) \left[ \left( 1 - { \bar{Z} \over x_\p} \right) r \partial_r u^l + { 2\bar{Z} \over x_\p} u^l \right] \\
- 2(1-x_\p) u^l \Bigg\} \sin \theta \partial_\theta Y_l^m  \\
+ \bBp \Delta \Omega [ r \partial_rU^l - \bar{Z} r\partial_r u^l - 2(1-x_\p+\bar{Z})u^l ] \sin \theta \partial_\theta Y_l^m \\
+im\bB \Delta \Omega [ r\partial_r U^l - \bar{Z} r \partial_r u^l + 2(1-x_\p+\bar{Z})u^l] \cos \theta Y_l^m
\end{multline}

Finally, it is convenient to work with a ``divergence equation'' \cite{2fosc} that follows from the combination
\begin{multline}
\sin \theta \partial_\theta [ \sin\theta ( \rho_\n\mathcal{E}_\n^\theta + \rho_\p \mathcal{E}_\p^\theta )]
+ \partial_\varphi [  \rho_\n \mathcal{E}_\n^\varphi + \rho_\p \mathcal{E}_\p^\varphi] \quad \longrightarrow\\
-L[ x_\p \delta \Psi_\p^l + (1-x_\p)\delta \Psi_\n^l] Y_l^m
+2\Omega U^l [ L\cos \theta Y_l^m + \sin\theta \partial_\theta Y_l^m ] = \\
= - \Delta \Omega (1-x_\p) \left\{ 
2U^l[ L\cos \theta Y_l^m + \sin\theta \partial_\theta Y_l^m ] 
-Lx_\p u^l[ 2\cos \theta Y_l^m + \sin\theta \partial_\theta Y_l^m ]  \right. \\
\left. + (L-2)\bar{Z}u^l  \sin\theta \partial_\theta Y_l^m \right\}
\end{multline}

Analogously, we consider the ``difference'' equation;
\begin{multline}
\sin \theta \partial_\theta [ \sin\theta ( \mathcal{E}_\p^\theta -  \mathcal{E}_\n^\theta )]
+ \partial_\varphi [  \mathcal{E}_\p^\varphi - \mathcal{E}_\n^\varphi] \quad \longrightarrow\\
L(\delta\Psi_\n - \delta \Psi_\p) Y_l^m + 2\Omega (1-\bBp) u^l [ L\cos \theta Y_l^m + \sin\theta \partial_\theta Y_l^m ] \\
+ 2im\Omega \bB u^l [ 2\cos \theta Y_l^m + \sin\theta \partial_\theta Y_l^m ] = \\
= \Delta \Omega L ( U^l - x_\p u^l) [ 2\cos \theta Y_l^m + \sin\theta \partial_\theta Y_l^m ] \\
- \Delta \Omega \left[ {(L-2) \bar{Z} \over x_\p} 
(U^l +u^l) - (L-2)(1-x_\p + 2\bar{Z})  u^l \right]\sin\theta \partial_\theta Y_l^m \\
+\Delta \Omega \bBp \left\{ -LU^l + [2(Z+x_\p)-(1-x_\p)]Lu^l \right\} [ 2\cos \theta Y_l^m + \sin\theta \partial_\theta Y_l^m ]\\
- \Delta \Omega \bBp (L-2)(Z+x_\p) u^l \sin\theta \partial_\theta Y_l^m  \\
-2im\Omega \bB \Delta (x_\p + Z) u^l [ 2\cos \theta Y_l^m + \sin\theta \partial_\theta Y_l^m ] \\
-im\Omega \bB \Delta [ 2 r \partial_r U^l + LU^l + (1-2x_\p-Z)(2r \partial_r u^l +Lu^l)]\cos \theta Y_l^m
\end{multline}

Next, we separate the $l$-multipoles by means of the standard recurrence relations
\be
\cos \theta Y_l^m = Q_{l+1} Y_{l+1}^m + Q_l  Y_{l-1}^m
\ee
and
\be
\sin \theta \partial_\theta Y_l^m = l Q_{l+1} Y_{l+1}^m - (l+1) Q_l  Y_{l-1}^m
\ee
where
\be
Q_l^2 ={ (l-m)(l+m)\over (2l-1)(2l+1)}
\ee
These relations lead to 
\be
\cos^2 \theta Y_l^m = ( Q_{l+1}^2 + Q_l^2) Y_l^m + Q_{l+1} Q_{l+2}Y_{l+2}^m
+Q_l Q_{l-1} Y_{l-1}^m
\ee
and
\be
\cos\theta \sin \theta \partial_\theta Y_l^m = [lQ_{l+1}^2 - (l+1)Q_l^2] Y_l^m 
+ l Q_{l+1} Q_{l+2} Y_{l+2}^m - (l+1) Q_l Q_{l-1}Y_{l-2}^m 
\ee

It should be quite clear at this point that the problem we consider is rather complex, even for purely axial modes. One must also 
be careful, because it is not clear from the outset that modes of this particular character exists (as, in principle, rotation would couple the axial/toroidal degree of freedom to the polar/spheroidal one \cite{lock}). However, in the particular case that we are considering the problem
simplifies in an almost miraculous fashion. We do not expect this level of simplification in a more general situation, leaving the  
problem exceedingly difficult. 


\section{The unstable r-modes}

The r-modes are very special members of the general class of inertial modes, because their eigenfunctions ``truncate'' at $l=m$ (at least in the 
standard single-fluid setting) making their eigenfunctions particularly simple (generic inertial modes involve coupling a number of $l\ge m$ multipoles \cite{lock}).
Inspired by the fact that this remains true also for co-rotating superfluids \cite{2fosc}, it makes sense to ask whether it may be the case when a rotational lag is present as well. Somewhat to our surprise, it turns out that such simple r-mode solutions do, indeed, exist. Moreover, we find that these modes may become unstable once the vortex mediated mutual friction is accounted for.

Assuming that the perturbed velocity fields take the same form as in the co-rotating case, i.e. 
\be
U_{\rm x}^l = \left\{ \begin{array} {ll}  A_{\rm x} r^{m+1} \qquad \mbox{for } l = m \\
0  \qquad \mbox{for } l \ge m \end{array}\right.
\ee
and noting that $Q_{l=m}=0$, the equations from the previous section collapse to two scalar relations for the amplitudes. 
Expressing these in terms of $U^m$ and $u^m$, we have
\begin{multline}
[(m+1) \sigma - 2 + \Delta (1-x_\p) (m-1)(m+2)] U^m \\
- (1-x_\p - \varepsilon_\p)
\Delta x_\p (m-1) (m+2) u^m = 0
\label{ueq1}\end{multline}
and
\begin{multline}
- \left[ (m-1)(m+2) +  2 \bep  - m(m+1) (\bBp+i\bB) \right]\Delta U^m \\
+ \Big\{ (1-\bep) (m+1)\sigma - 2(1-\bBp+i\bB) +\Delta x_\p (m-1)(m+2) \\
- \bep \Delta \left\{ [ m(m+1)-4]x_\p + 2 \right\}
- m(m+1) \Delta x_\p (1-\bep) ( \bBp + i\bB)
+ 2(1-\varepsilon_\n) (\bBp - i\bB) \Delta
 \Big\} u^m
=0
\label{ueq2}\end{multline}
where we have used $\bar{\varepsilon} = \varepsilon_\n/x_\p = \varepsilon_\p/(1-x_\p)$. In the present case, where all the coefficients are taken to be constant, it is easy to see that the problem reduces to a quadratic for $\sigma$. The general solution to this quadratic is, however, rather messy 
and not particularly instructive. In order to understand the nature of the solutions, it is better to focus on simplified cases.

\subsection{Dynamical instability}

Let us first consider the case of vanishing entrainment.  In this case we have
\be
(m+1)^2 \sigma^2 + P_1 \sigma + P_0 = 0
\label{dispo}\ee
with
\begin{multline}
P_0 = 2(m-1)(m+2) \left( \bar {\cal B}^\prime -i \bar{\cal B} \right ) \Delta^2
+ 4 \left [ 1 - \left (1 +  x_{\rm p} \right )  \left ( \bar{\cal B}^\prime -i \bar {\cal B} \right ) \right ]
\\ 
+ 2 \Delta \left [ (m-1)(m+2) (x_\p \bar {\cal B}^\prime -1 ) +   \left ( \bar {\cal B}^\prime -i \bar {\cal B} \right )
(m^2 + m -4) + ix_\p \bar {\cal B} (m^2 + m +2) \right ]
\end{multline}
and
\begin{multline}
 P_1 = (m+1) \Big\{  -4+ 2 \left(1+x_\p\right) \left( \bar {\cal B}^\prime -i \bar {\cal B} \right)\\
 + \Delta \left[ 2
\left ( \bar {\cal B}^\prime -i \bar {\cal B} \right ) + (m-1)(m+2) (1- x_\p \bar {\cal B}^\prime ) -i x_\p \bar {\cal B}
(m^2 + m +2) \right] \Big\}
\end{multline}

\begin{figure}[h]
\centerline{\includegraphics[height= 8cm,clip]{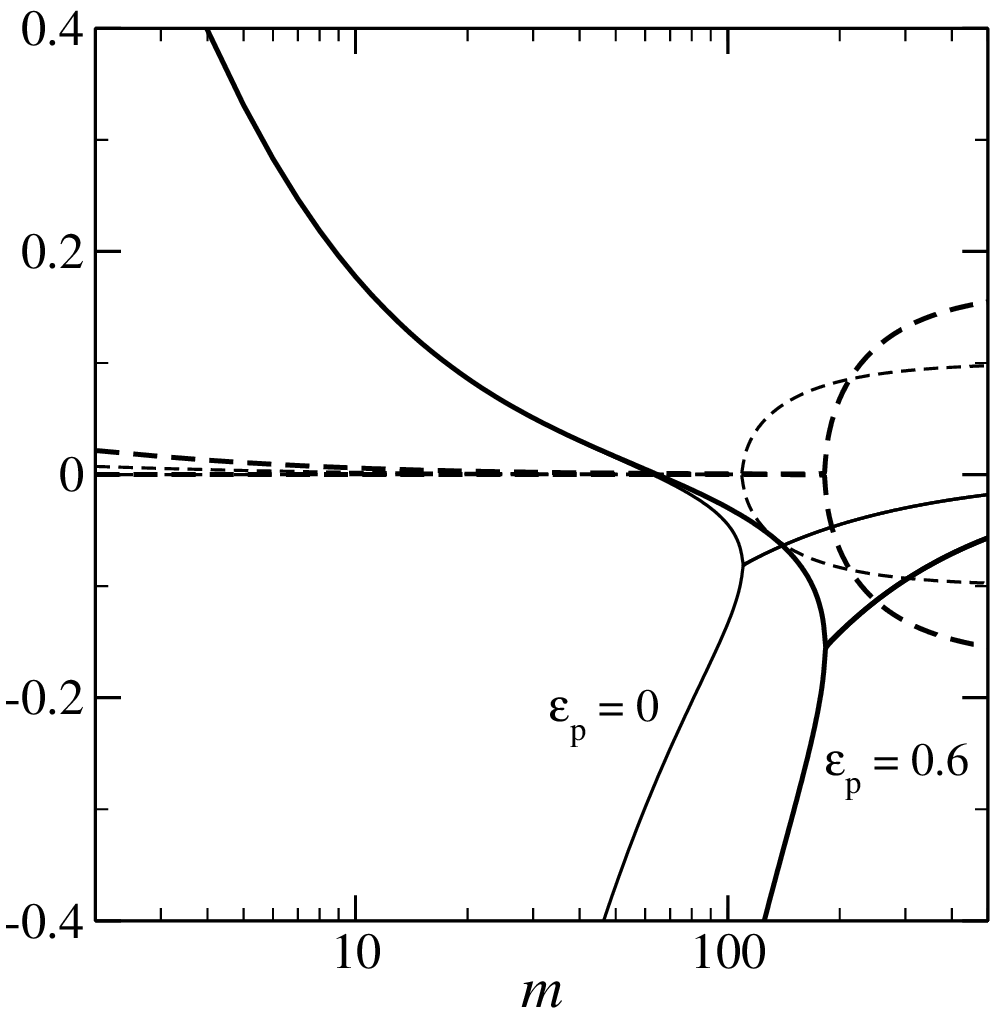}}
\caption{Real (solid lines) and imaginary parts (dashed) of the roots to the dispersion relation (the r-modes) in the strong-coupling limit, $\mathcal R = 10^3$, for two illustrative cases, when $\varepsilon_\p=0$ (thin lines) and $\varepsilon_\p=0.6$ (thick lines). The rotational lag is fixed to $\Delta = 5\times 10^{-4}$. In each case, the presence of a critical value for the azimuthal index $m_c$, beyond which the modes are unstable (the roots are complex) is apparent. The results also demonstrate how the entrainment affects the range of the instability by shifting $m_c$. Meanwhile, the growth rate of the instability (the magnitude of the imaginary part in the unstable regime) is not affected much.  }
\label{fig2}
\end{figure}

The explicit solutions are, of course, still not transparent. However, if we consider the limit  
of strong mutual friction coupling ($\mathcal R\to \infty$), such that ${\cal B}\approx 0$ and ${\cal B}^\prime \approx 1 $, then 
we obtain the roots
\be
\sigma= -\frac{1}{(m+1) x_\p} \left [ 1 -x_\p + \Delta \pm {\cal D}^{1/2} \right ]
\label{strongroots}
\ee 
with 
\be
{\cal D} = (1+x_\p)^2 + 2\Delta \left\{ 1 + x_\p \left[ 3 -m(m+1) \right]\right\}
\ee
These solutions highlight one of the main new results in this paper (and \cite{GAprl}): 
For $ m \gg 1  $ we  have 
\be
{\cal D} \approx (1+x_\p)^2 - 2 x_\p m^2 \Delta \ , 
\ee
 showing that 
we have  {\it unstable} r-modes (${\rm Im}\ \sigma<0 $)  for
\be
m \gtrsim m_c \ , \qquad \mbox{where} \qquad  \quad m_c = \frac{1+x_\p}{\sqrt{2x_\p \Delta}}
\label{mc}
\ee
Expressed in terms of the (observed) rotation period $P=2\pi/\Omega$ of the system, 
the growth timescale for these unstable modes is
\be
\tau_{\rm grow} \approx \frac{m P}{2\pi} \left ( \frac{x_\p}{1+x_\p} \right ) \left (\frac{m^2}{m^2_c} -1 \right )^{-1/2}
\ee 
For  $m \gg m_c $ (in practice, $m \gtrsim 2 m_c $), the growth rate is well approximated by
\be
\tau_{\rm grow} \approx \frac{P}{2\pi} \left ( \frac{x_\p}{2\Delta} \right )^{1/2} \quad
\quad \Rightarrow  \quad \tau_{\rm grow} \approx 3 \left ( \frac{x_\p}{0.05} \right )^{1/2}
\left ( \frac{10^{-4}}{\Delta} \right )^{1/2}\, P
\label{grow}
\ee
With the same scaling, the critical multipole beyond which the instability is present is
\be
m_c \approx 300 \left( {x_\p \over 0.05} \right)^{-1/2} \left( {\Delta \over 10^{-4} }\right)^{-1/2}
\ee
The presence of this critical value, and the associated emergence of unstable modes, is illustrated in Figure~\ref{fig2}. We see that the instability sets in as two r-modes merge at a critical scale represented by $m_c$. This is the characteristic behaviour of a dynamical instability in a non-dissipative system \cite{gwrev}.

Accounting for the entrainment obviously complicates the analysis. However, the result remains transparent in the strong coupling limit.
Again setting ${\cal B} =0$ and ${\cal B}^\prime =1 $, we find the roots
\be
\sigma = \frac{\gamma}{(m+1) x_\p} \left [ -(1+\varepsilon_\n) + (1-\varepsilon_\n)( x_\p -\Delta ) \pm {\cal D}^{1/2} \right ] 
\ee
with
\be
\gamma = (1-\varepsilon_\n-\varepsilon_\p)^{-1}
\ee
and
\begin{multline}
{\cal D} = \left ( 1 + \frac{x_\p}{\gamma} \right )^2 + 2(1-\varepsilon_\n) \left [ 1- (m^2 +m -3)\frac{x_\p}{\gamma}  \right ]\Delta \\
+ (1-\varepsilon_\n)^2 \left [ 1 - \frac{2x_\p}{\gamma} (m+2) (m-1) \right ] \Delta^2 
\end{multline}
For $m \gg 1 $ and $\Delta \ll 1 $ this is approximately,
\be
{\cal D} \approx \frac{(\gamma + x_\p )^2}{\gamma^2} \left ( 1 -\frac{m^2}{m^2_c} \right ), \qquad \mbox{where} \qquad
m_c^2  = \frac{(\gamma + x_\p)^2}{2x_\p \Delta \gamma (1-\varepsilon_\n)} 
\ee
We see that, in this case there are unstable modes with growth time;
\be
\tau_{\rm grow} \approx 
\frac{m P}{2\pi} \left ( \frac{x_\p}{\gamma +x_\p} \right ) \left (\frac{m^2}{m^2_c} -1 \right )^{-1/2}
\label{tgrow1}
\ee
For $ m \gg m_c $ this reduces to
\be
\tau_{\rm grow} \approx  \frac{P}{2\pi} \left ( \frac{x_\p}{2\Delta} \right )^{1/2} \epsilon_\star^{1/2}
\ee
where we have  introduced 
\be
\epsilon_\star = \frac{1-\varepsilon_\n -\varepsilon_\p}{1-\varepsilon_\n} \approx \frac{m_\p^*}{m_\p}
\ee
Given that a typical range of values for the effective proton mass, $m_\p^*$, in a neutron star core is expected to be in the range  $m_\p^*/m_\p \approx 0.5 -0.9$ \cite{slowrot}
we conclude that entrainment has a minor impact on the growth timescale. An illustration of this result is provided in Figure~\ref{fig2}.

In summary, these results show that, for the typical magnitude of rotational lag inferred from radio pulsar glitches \cite{glitch} we would have unstable modes with a characteristic horizontal length-scale of tens to hundreds of metres.
Smaller scale r-modes would be unstable, with a growth time as fast as a few rotation periods. As the unstable modes
grow extremely fast compared to the evolutionary timescale (spin-down, cooling etc) of the system it seems reasonable to
expect that they may affect  any real system that develops the required lag, $\Delta$. It is worth noting that the predicted growth time is much shorter than the 
current observational constraint for the rise of pulsar glitches (tens of seconds \cite{velacon}), unless the star is slowly rotating. Hence, the instability could grow fast enough to serve as trigger for the observed events.

\subsection{Secular instability}

\begin{figure}[h]
  \includegraphics[width=6in,clip]{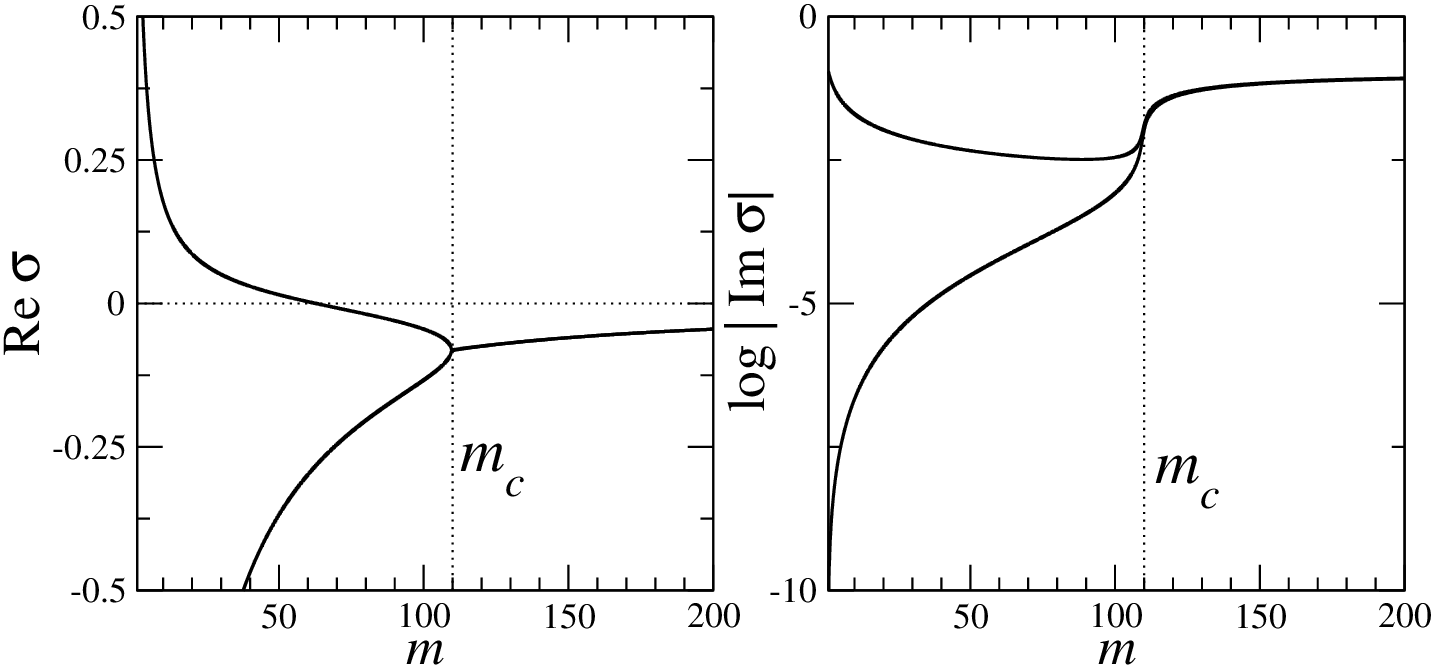}
  \caption{Real (left panel) and imaginary parts (right panel) of the r-modes for $\mathcal R=100$, $x_\p=0.1$ and a rotational lag $\Delta = 5\times10^{-4}$.  Beyond a critical critical value for the azimuthal index, $m_c$, the modes are dynamically unstable. The onset of this instability is associated with (near) merger of the real part of the two frequencies. The absence of sign change of the imaginary part of the unstable branch shows the presence of a secular instability for smaller values of $m$. }
\label{mfig100}\end{figure}

\begin{figure}[h]
  \includegraphics[width=6in,clip]{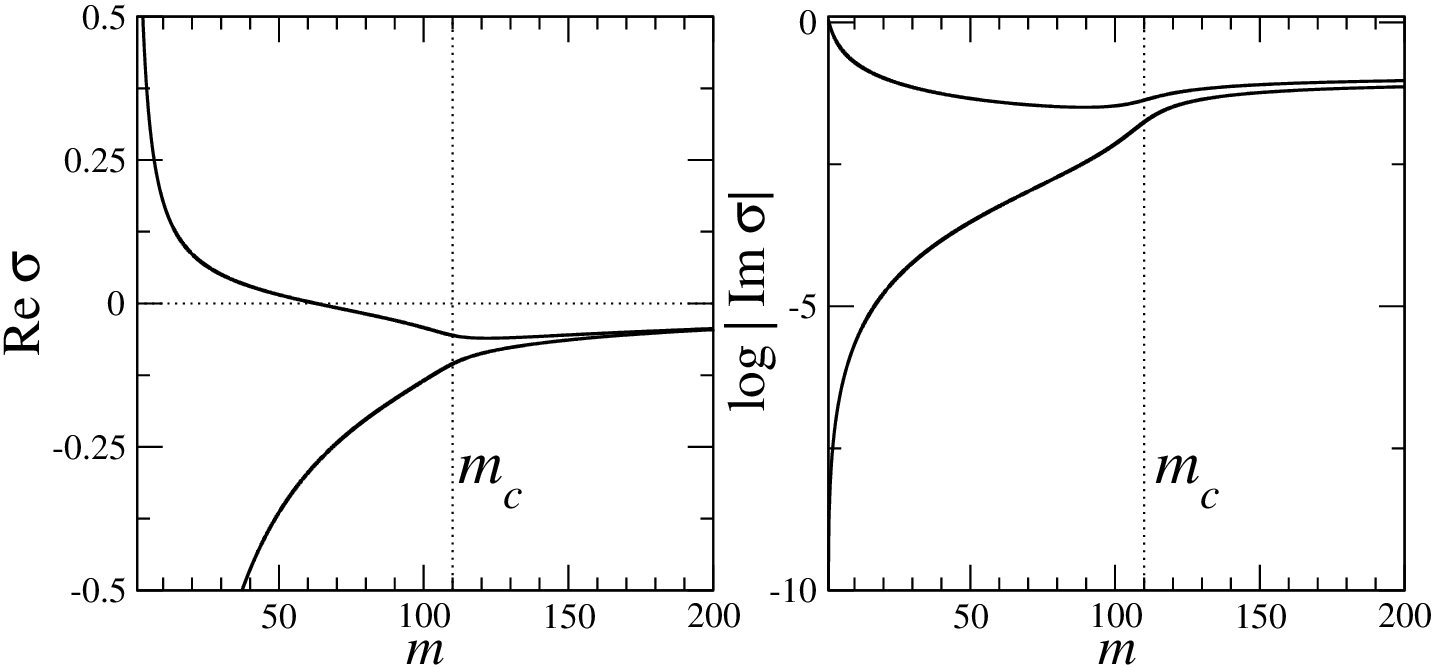}
  \caption{Same as Fig.~\ref{mfig100} but for  ${\mathcal R} =10$.}
\label{mfig10}\end{figure}

\begin{figure}[h]
  \includegraphics[width=6in,clip]{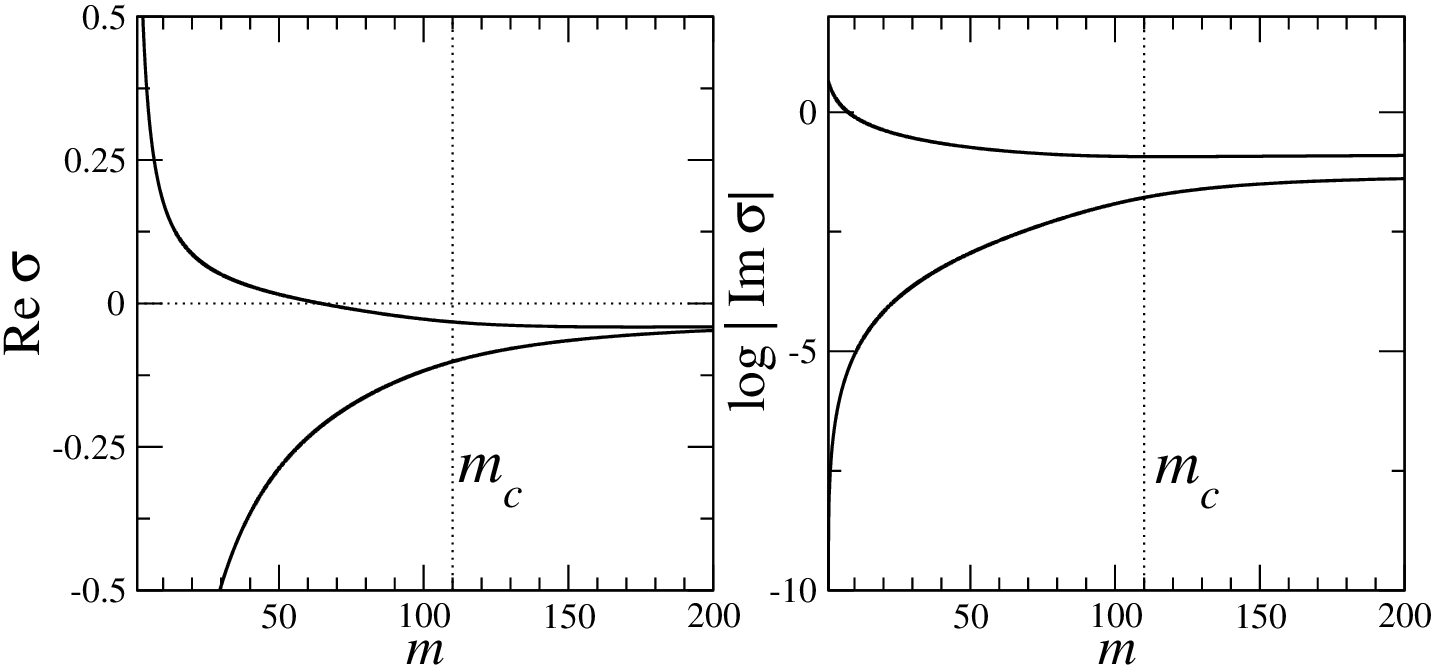}
  \caption{Same as Fig.~\ref{mfig100} but for ${\mathcal R} =2$.}
\label{mfig2}\end{figure}

Having established the existence of an instability in the strong-coupling limit, let us  consider the problem for less
``extreme'' parameters. It is, of course, straightforward to solve \eqref{dispo} for given parameter values. A sample of results obtained by considering the problem for
fixed $\mathcal R$, leading to the mode frequencies depending on $m$, are provided in Figs.~\ref{mfig100}--\ref{mfig2}. These graphs show the behaviour of the r-mode solutions as $\mathcal R$ decreases from 100 to 2, i.e. as we move away from the strong-coupling regime. The results are very interesting. First of all, we see that the general trend of a ``dynamical'' instability (associated with modes with a markedly larger imaginary part to their frequency) setting in near the critical value $m_c$ remains down to $\mathcal R=10$ or so. For smaller values, e.g. $\mathcal R=2$ as in Fig.~\ref{mfig2}, there is no longer a clear change in the imaginary part near $m_c$.
We also see that, away from the extreme strong coupling limit the dissipative aspect of the mutual friction becomes important. In particular, it leads to the dynamical instability no longer being associated with exact mode mergers. Rather, the instability sets in at "near misses" in the complex frequency plane. This is probably what should be expected. Another aspect of the results was  not expected. Considering the imaginary parts of the roots in more detail (the right-hand panels of Figs.~\ref{mfig100}--\ref{mfig2}) we see that neither imaginary part changes sign in the displayed interval (as we show $\log | \mathrm{Im}\ \sigma|$ a sign change would show up as a sharp singularity). This demonstrates the most interesting new result in this paper. As we know that one of the modes is unstable beyond $m_c$ we must conclude from Figs.~\ref{mfig100}--\ref{mfig2} that this mode is, in fact, unstable for all lower values of $m$, as well. Of course, for smaller $m$ the growth time of the unstable modes is much longer. Later we will demonstrate that it is linked to the mutual friction parameters, making this a secular instability (plausibly related to the Donnelly-Glaberson vortex instability in laboratory superfluids \cite{dg1,dg2,sid2}).
Particularly interesting may be the fact that this instability is not restricted to the small scale r-modes. It is also active for the large scale modes. This is interesting since these modes, expecially the $m=2$ r-mode, are also secularly unstable due to gravitational-wave emission. Basically, our results show that the mutual friction may not provide damping of these modes. Rather, it could provide an additional driving mechanism for the instability. It may also be, given the strong scaling of the gravitational radiation reaction with the modes oscillation frequency (essentially the star's spin rate) that the mutual friction driven instability dominates for slowly rotating systems. We will return to this question later. 

Let us first see if we can find approximate solutions that demonstrate the behaviour seen in the numerical results.
Returning to \eqref{ueq1} and \eqref{ueq2} we see that the  two degrees of freedom
(represented by $U^m$ and $u^m$) only couple at order $\Delta$. Since we are  interested in mode solutions for small $\Delta$, this 
suggests that the solutions are either predominantly co- or counter-moving, depending on whether  $U^m$ dominates over $u^m$, or vice versa. 
Considering first the co-moving case, we assume that  $U^m\neq0$ and $u^m\approx 0$, which leads to
\be
\sigma= { 1 \over m+1} \left[ 2 - \Delta (1-x_\p) (m-1)(m+2) \right] \equiv\sigma_0
\label{app1}\ee
At this level, the frequency is real and the modes are stable. 

Meanwhile, in the counter-moving case we have modes with frequency
\begin{multline}
\sigma = { 1 \over (1-\bep) (m+1)} \Big\{2(1-\bBp+i\bB) -\Delta x_\p (m-1)(m+2)
+ \bep \Delta \left\{ [ m(m+1)-4]x_\p + 2 \right\} \\
+ m(m+1) \Delta x_\p (1-\bep) ( \bBp + i\bB)
- 2(1-\varepsilon_\n) (\bBp - i\bB) \Delta  \Big\}
\end{multline}
From this we see that the imaginary part of the frequency is proportional to
\begin{displaymath}
2 + m(m+1) \Delta x_\p (1-\bep) + 2(1-\varepsilon_\n) \Delta
\label{app2}\end{displaymath}
which suggests that the counter-moving modes tend to be stable, at least in the non-entrainment case.
When the entrainment is accounted for, these modes may become unstable. In the  short lengthscale limit, when $m^2 x_\p \Delta \gg 1$ (corresponding to $m\gg 10^3$ or so for typical parameters), an instability is present as long as $\varepsilon_\n > x_\p$, which is not a particularly severe criterion.

Let us now consider the co-rotating modes at the next level of approximation. We have already seen from \eqref{app1} that, 
to leading order in $\Delta$, these modes are marginally stable,
with frequency approximated by $\sigma_0$.  Yet, we know from the numerical results that an instability should be present for some range of $\mathcal R$, certainly $\mathcal R\ge 2$. Hence, we need to  estimate the solution at the next order in  $\Delta$. To do this, we take
\be
\sigma = \sigma_0 + \sigma_2 \Delta^2
\ee
which leads to
\be
\sigma_2  = {(m-1) (m+2)\over 2 (m+1)} { x_\p (1-x_\p) \over (\bBp-i\bB)}
 \left[  (m-1)(m+2) - m(m+1) (\bBp+i\bB) \right]
\label{imap2}\ee
We are (primarily) interested in the imaginary part of the frequency. Since we know that an instability exists in the 
strong-coupling limit we note that this limit corresponds to  $\bBp\approx 1/x_\p$ and hence we have (to order $\bB$)
\be
\mathrm{Im}\ \sigma_2 = -{(m-1) (m+2)\over 2 (m+1)} x_\p (1-x_\p) [m(m+1)(2-x_\p)+2x_\p] \bB
\label{app3}\ee
We learn that these modes are {\em unstable} for {\em all} values of $\Delta$. We also see that the growth rate scales as $1/\bB$ making this a secular instability. The estimate \eqref{app3} establishes the presence of 
the instability for a wide range of parameters, and  provides more detailed insight into the dependence on the parameters.

It is obviously relevant to establish what the critical value of $\mathcal R$ may be. We can get an idea of this by working out where the imaginary part of \eqref{imap2} changes sign. This leads to
\be
\mathcal B'_c = {\mathcal R_c^2 \over 1 + \mathcal R_c^2} =  { (m-1) (m+2) \over 2m (m+1)} x_\p
\label{Rcrit}\ee
For  $\mathcal R\le \mathcal R_c$ the system should be stable. It is easy to show that this is the case by considering the weak-coupling limit. Then we have $\bBp \sim \bB^2$ so the dominant behaviour will be
\be
\mathrm{Im}\ \sigma_2 = {(m-1)^2 (m+2)^2\over 2 (m+1)} x_\p (1-x_\p) {1 \over \bB}
\label{app4}\ee
As expected, the modes are {\em always stable} in this limit. This is, of course, as expected. 

\begin{figure}
  \includegraphics[width=6in,clip]{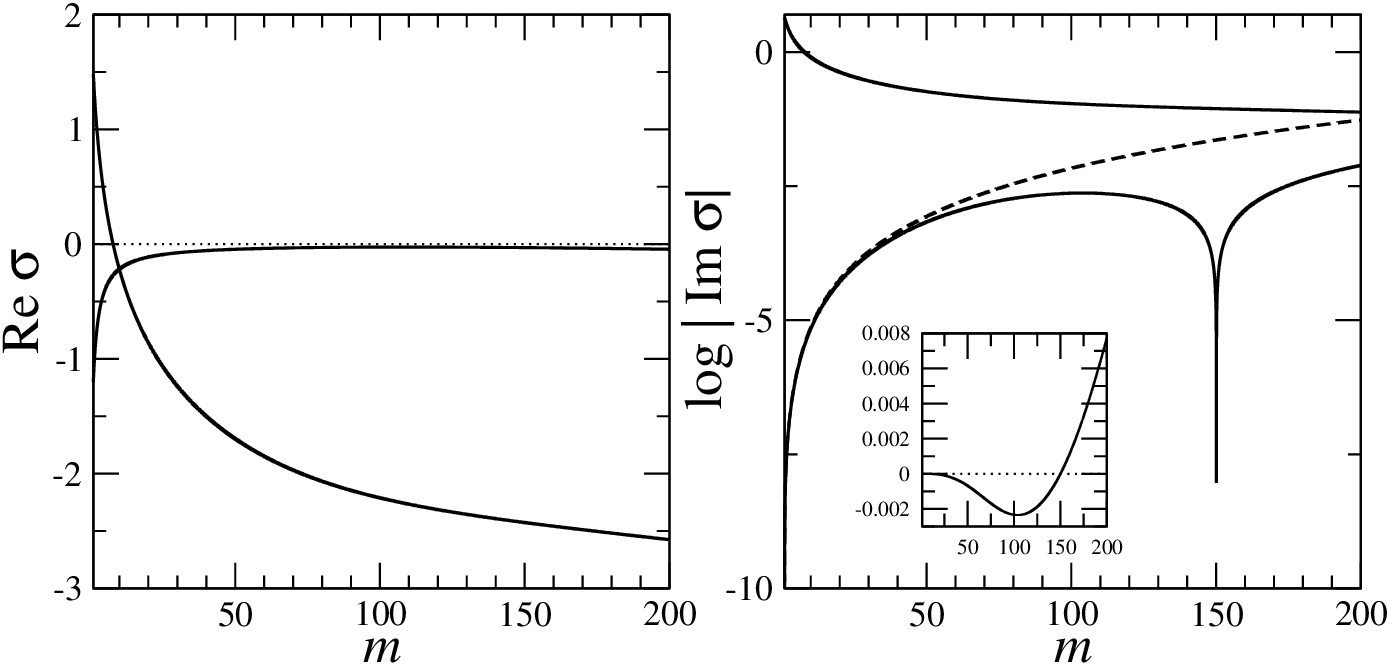}
  \caption{Same as Fig.~\ref{mfig100} but for ${\mathcal R} =0.5$. The inset in the right panel shows the detailed behaviour of  $\mathrm{Im}\ \sigma$ as function of $m$. The dashed curve in the right-hand panel corresponds to the approximate solution \eqref{imap2}, which is an accurate representation for low multipoles.}
\label{mfig05}\end{figure}

Returning to the numerical results, the typical behaviour for $ \mathcal R_c < \mathcal R < 1$ is similar to that shown in Fig.~\ref{mfig05} (which corresponds to $\mathcal R=0.5$). That is, for a given value of $\mathcal R$ the instability is present for a range of multipoles, up to a critical value where the modes become stable. Above $\mathcal R = 1$ all modes are unstable.

\begin{figure}[h]
     \includegraphics[height=6cm,clip]{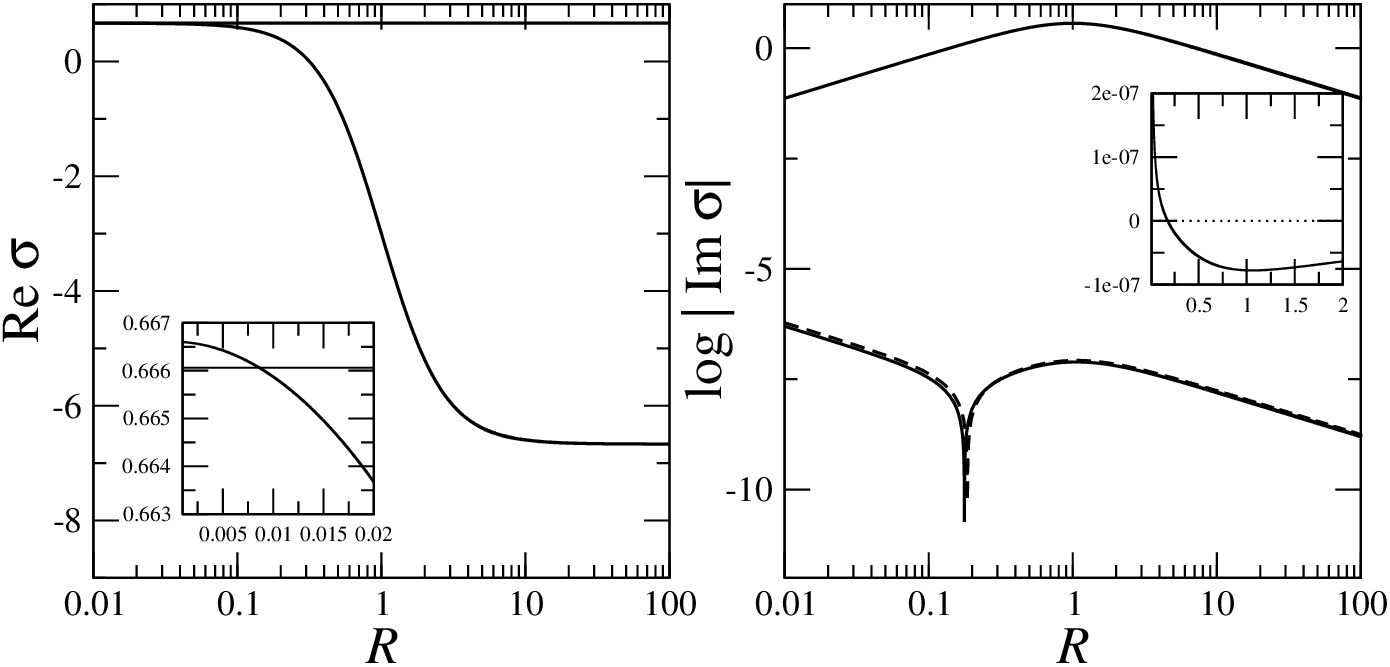}
     \caption{Real (left panel) and imaginary parts (right panel) of the r-modes for $m=2$, $x_\p=0.1$ and a rotational lag $\Delta = 5\times10^{-4}$. Unstable modes are present beyond a critical value of $\mathcal R$, see inset in the right panel. We compare the obtained imaginary part for the  modes that become unstable to the estimate
\eqref{imap2} (dashed curve in the right-hand panel). In this case, this is clearly a very good approximation. The inset in the left panel shows that the real parts of the mode frequencies cross at a low value of $\mathcal R$, seemingly unrelated to the onset of instability.}
  \label{Rfig2}
\end{figure}

\begin{figure}[h]
     \includegraphics[height=6cm,clip]{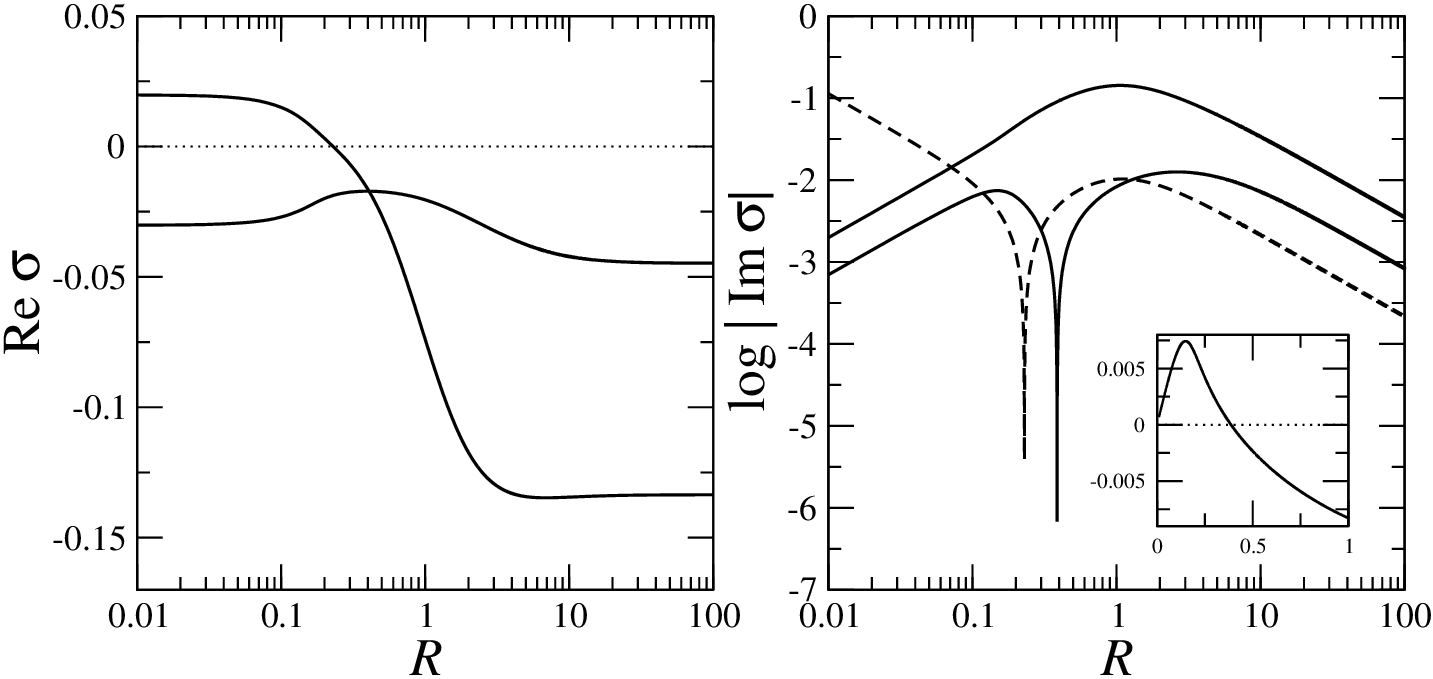}
     \caption{The same as Figure~\ref{Rfig2}, but for $m=100$. In this case, \eqref{imap2} no longer provides a good approximation to the imaginary part.}
  \label{Rfig100}
\end{figure}

We get a complementary view of the new instability by fixing the multipole $m$ and varying  $\mathcal R$. This leads to the results shown in Figures~\ref{Rfig2} and \ref{Rfig100}, for $m=2$ and $m=100$, respectively. These results demonstrate that the instability exists throughout the   $\mathcal R\ge 1$ regime, and  that is extends into the ``weak-coupling'' regime, as well. This is yet another demonstration of the generic nature of the  superfluid r-mode instability. It is notable that the approximation \eqref{app3} is excellent for low values of $m$, see the right-hand panel of Fig.~\ref{Rfig2}, but deteriorates as $m$ increases. This is not surprising since, for large values of $m$ one cannot simply neglect higher order terms in $\Delta$ as these may be multiplied by factors of $m$. In essence, the co- and counter-moving degrees of freedom are no longer neatly decoupled on shorter (angular) scales. Nevertheless, the approximation serves its purpose by illustrating the behaviour in an important part of parameter space.

\section{Astrophysical context: Accounting for shear viscosity}

As suggested in \cite{GAprl}, the superfluid instability discussed in the previous section could be relevant for
astrophysical neutron stars, in particular glitching pulsars. A ``minimum'' requirement for this to be the case is that
the instability  grows fast enough to overcome the dissipative action of viscosity in neutron star matter.

A mature neutron star core is sufficiently cold to contain both superfluid neutrons and protons (recent evidence suggest that this  is the case
for a core temperature $T \lesssim 5-9\times10^8\,\mbox{K}$ \cite{casa,casb}. Under these conditions the fluid motion is primarily damped by vortex mutual friction (which we have already accounted for, and which drives the instability we are considering) and shear viscosity due to electron-electron collisions  \cite{npa}. Drawing on the considerable amount of work that has been done on the gravitational-wave driven r-mode instability \cite{rmode}, we can estimate the shear viscosity damping timescale using an energy-integral approach;
\be
\tau_{\rm sv} = \frac{E_{\rm mode}}{\dot{E}_{\rm sv}}
\label{tsv1}
\ee
where $E_{\rm mode}$ is the mode energy and $\dot{E}_{\rm sv}$ is the shear viscosity damping rate.
The damping timescale (\ref{tsv1}) can be easily calculated using the two-fluid r-mode results of the previous sections.
However, as we are only interested in a rough estimate we use  the result for  ordinary single-fluid r-modes in a uniform density star \cite{ks} to approximate $\tau_{\rm sv}$. This leads to
\be
\tau_{\rm sv} = \frac{3}{4\pi} \frac{M}{\eta_{\rm ee} R} \frac{1}{(2m+3)(m-1)} 
\approx \frac{1.3 \times 10^6}{m^2} \left ( \frac{0.05}{x_\p} \right )^{3/2}
\frac{R_6^{7/2}}{M_{1.4}^{1/2}}T_8^2\, {\rm s} 
\label{tsv2}
\ee
where $R_6 = R/10^6\mbox{cm}$ and $M_{1.4} = M/1.4 M_\odot$ represent the radius and mass of the star, respectively.
The relevant shear viscosity coefficient ($\eta_{\rm ee}=\rho_\p \nu_{\e\e}$) has been taken to be \cite{npa}
\be
\eta_{\rm ee} = 1.5 \times 10^{19}\, \left (\frac{x_\p}{0.05} \right )^{3/2} \rho_{14}^{3/2} T_8^{-2}\,~
\mbox{g}\, \mbox{cm}^{-1}\, \mbox{s}^{-1}
\label{nee}
\ee
where $\rho_{14} = \rho/10^{14}\,\mbox{g}~\mbox{cm}^{-3}$ and $T_8 = T/10^8\,\mbox{K}$.

\begin{figure}
\centerline{\includegraphics[height= 8cm,clip]{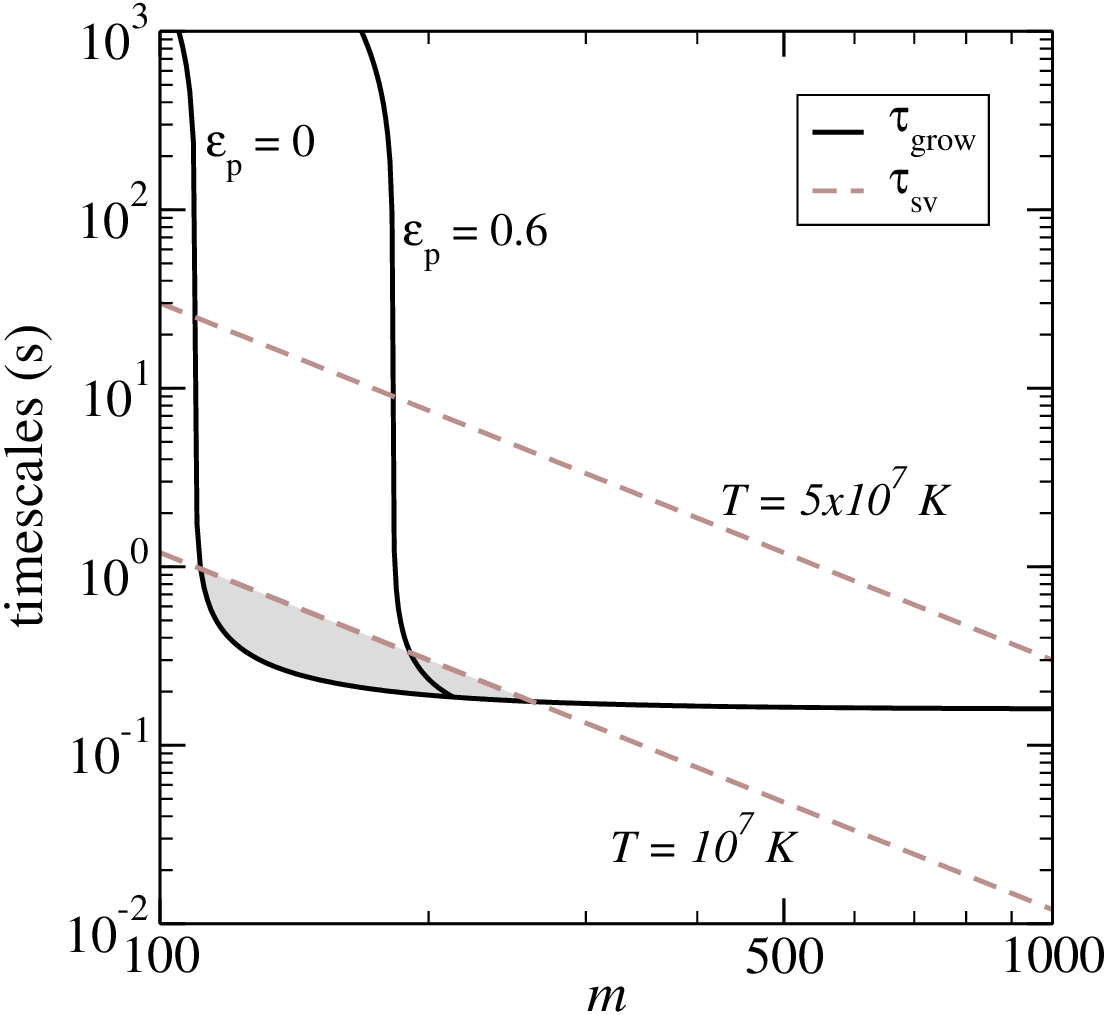}}
\caption{The dynamical r-mode instability growth and viscous damping timescales $\tau_{\rm grow}$ and $\tau_{\rm sv}$ (from Eqns.~(\ref{tgrow1}) and (\ref{tsv2}), respectively) as functions of the multipole $m$  for typical pulsar parameters: $P=0.1~\mbox{s},~x_\p=0.1$ and strong drag ${\cal R}=10^4$. The stellar mass and radius are fixed at the canonical values $M_{1.4} = R_6 =1$. The shear viscosity damping rate $\tau_{\rm sv}$ is shown for two core temperatures: $T=10^7\,~\mbox{K}$ and $T=5\times 10^7\,~\mbox{K}$.
The spin-lag $\Delta$  is fixed to $5\times10^{-4}$ and we show results for two values of the entrainment parameter, $\varepsilon_\p = 0$ and $\varepsilon_\p = 0.6$. The grey region indicates the range of unstable r-modes for  $T=10^7\,~\mbox{K}$. }
\label{fig3}
\end{figure}

We are now in a position where we can compare the growth timescale, $\tau_{\rm grow}$, to that due to shear-viscosity damping, $\tau_{\rm sv}$. Such a comparison is provided in Figure~\ref{fig3} for modes exhibiting the dynamical instability (in the strong-coupling regime) and  typical neutron-star parameters. The timescales are shown as functions of the multipole $m$, and we provide results for  two choices of 
the entrainment parameter $\varepsilon_\p$, cf. Fig.~\ref{fig2}. 
The damping rate $\tau_{\rm sv}$ is shown for two representative core temperatures $T=10^7\,\mbox{K}$ and  $5 \times 10^7\,\mbox{K}$, representing mature neutron stars. 
Dynamically unstable superfluid r-modes exist for the range of $m$ {\em above} the $\tau_\mathrm{grow}$ curve but {\em below} the $\tau_\mathrm{sv}$ curve. The results show that the $\tau_{\rm grow}$ profile levels off (as a function of $m$) for $m \gtrsim m_c$ and
that the instability can overcome the viscous damping for a range of scales.  We also see that the entrainment has a small effect on the asymptotic behaviour of $\tau_{\rm grow}$ but can significantly 
affect the critical multipole $m_c$. Finally, it is apparent that the range of unstable modes  decreases as the star cools. This is an interesting observation since glitches are only seen in relatively young pulsars. The results in Figure~\ref{fig3} indicate that shear viscosity would prevent the instability from developing soon after the star has cooled below $10^7$~K. It is  worth keeping in mind that the core temperature may remain above $10^8$~K for (at least) the first $10^5$ years of a pulsar's life \cite{homag}.

The predicted spin-lag for the dynamical instability to set in also makes a connection with pulsar glitches seem plausible. Balancing
the mode growth and the viscous damping, i.e., setting $\tau_{\rm grow} =\tau_{\rm sv} $, we find the critical spin lag $\Delta_c$
above which the instability is active. Combining (\ref{grow}) and (\ref{tsv2}) and setting $m =m_c$ we obtain 
\be
\Delta_c \approx 3.3 \times 10^{-5}  \left ( \frac{x_\p}{0.05} \right )^{2/3} \left (\frac{P}{1\,\mbox{s}}\right)^{2/3}
\left ( \frac{m^*_\p}{m_\p} \right )^{-1/3}
T_8^{-4/3} 
\label{lag_crit}
\ee
This result was first derived in  \cite{GAprl}. As discussed in that paper, the predicted critical lag compares well 
with the available data for large pulsar glitches. This could be an indication that large glitches
are indeed triggered by the large $m$ r-mode instability. 

\begin{figure}
{\includegraphics[height= 6cm,clip]{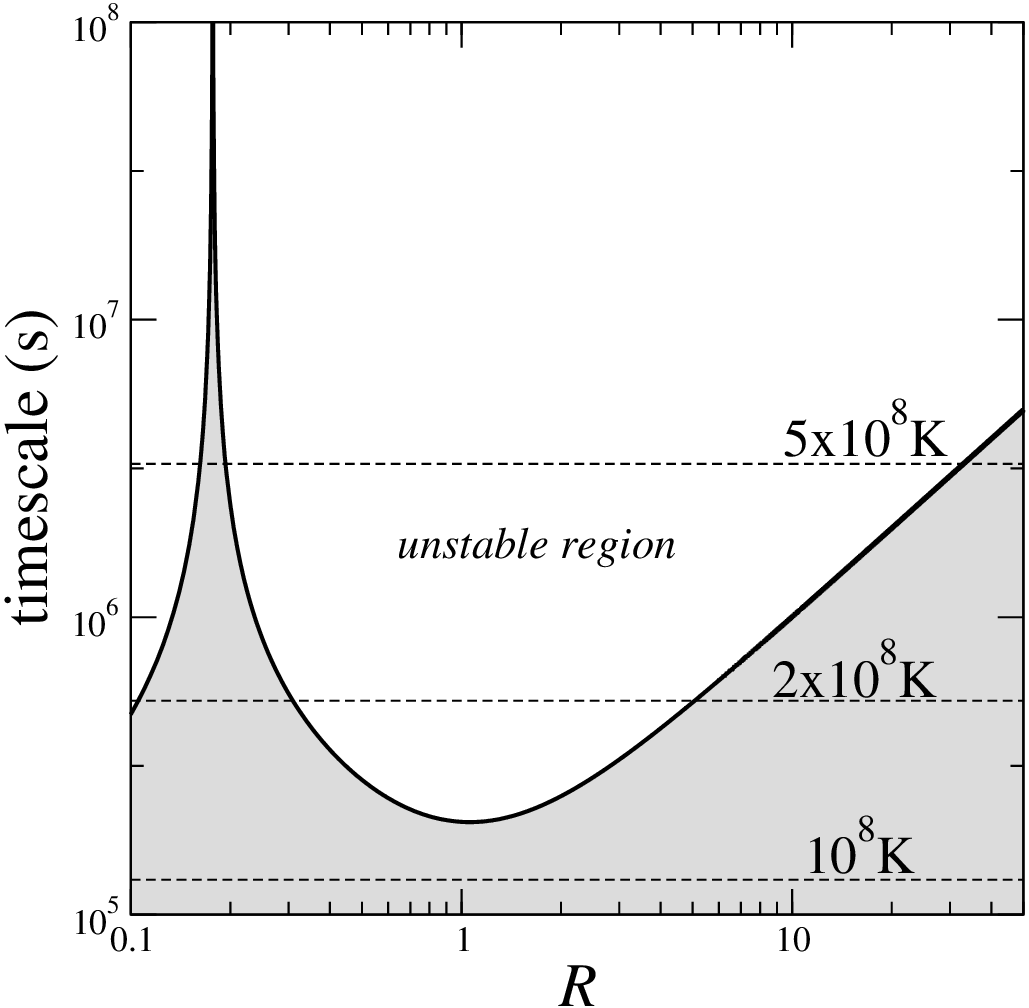}}{\includegraphics[height= 6cm,clip]{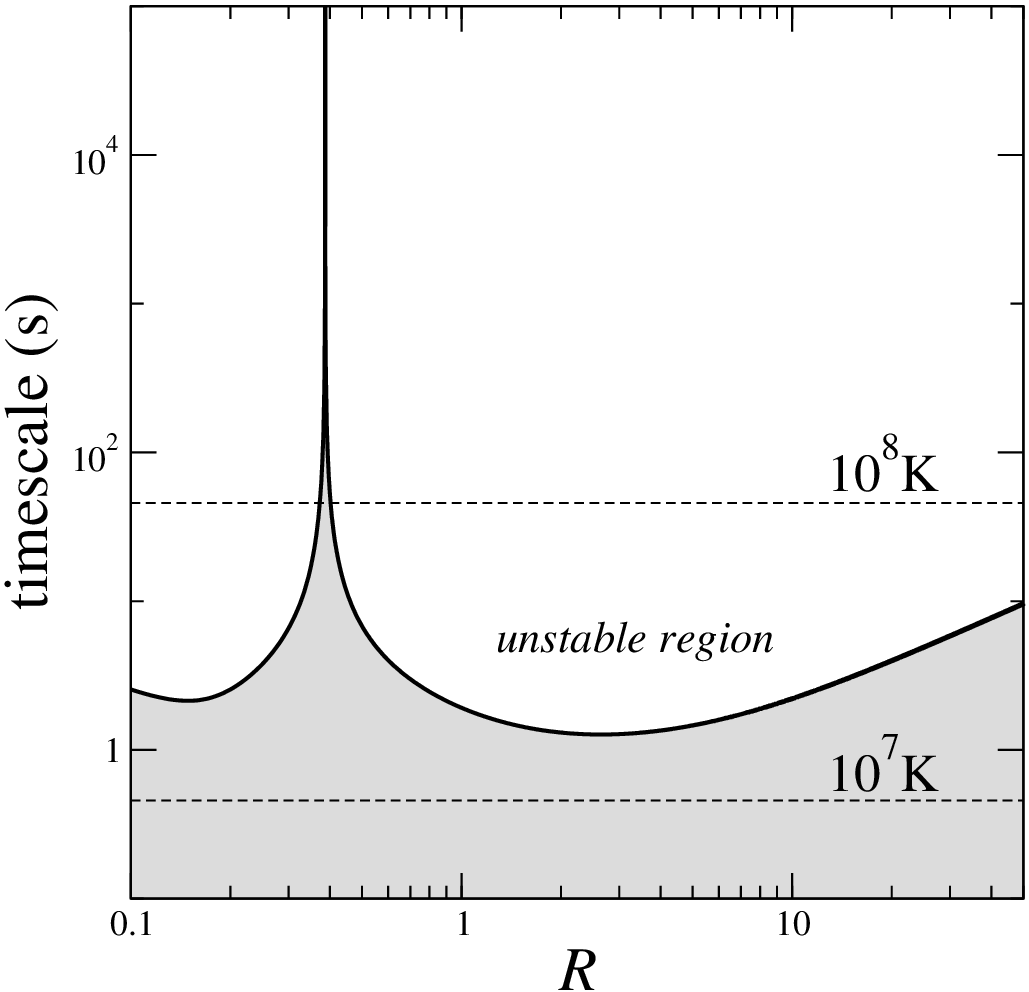}}
\caption{The secular r-mode instability growth and viscous damping timescales $\tau_{\rm grow}$ and $\tau_{\rm sv}$ (obtained from the numerical solution to \eqref{dispo} and (\ref{tsv2}), respectively) as functions of $\mathcal R$ for two multipoles: $m=2$ (left panel) and $m=100$ (right panel). The other parameters are the same as in Fig.~\ref{fig3}: $P=0.1~\mbox{s}$,  $x_\p=0.1$ and $M_{1.4} = R_6 =1$. The spin-lag $\Delta$  is fixed to $5\times10^{-4}$. The shear viscosity damping rate $\tau_{\rm sv}$ is shown for various temperatures as indicated in each panel (dashed horizontal lines). The r-mode growth time is short enough for an instability to be present for a range of $\mathcal R$ for temperatures characteristic of young neutron stars.}
\label{sectimes}
\end{figure}

The results for the secular instability for smaller values of $m$ are quite similar. In Fig.~\ref{sectimes} we show results both for $m=2$ and $m=100$. In each case we see that unstable modes will be present above a certain temperature. Keeping in mind that the critical temperature for core superfluidity is expected to be below $10^9$~K, while a typical glitching pulsar like the Vela is expected to have core temperature just above $10^8$~K, these results make it seem plausible that the secular instability can become active in these systems. Whether astrophysical systems with $\mathcal R$ in the required range for the instability exist is not clear at this point. Most work has focussed on smaller values, in which case the system would not exhibit the instability we have discussed here, but there have been suggestions of larger values \cite{sedrak}. It is also worth noting that a stronger mutual friction  may  help reconcile the discrepancy between our theoretical understanding of the r-mode instability with observed astrophysical systems \cite{hoetal}.


\section{Final Remarks}

We have provided a more detailed analysis of the superfluid r-modes in a simple uniform density model with (solid body) differential rotation between the two components. As already discussed in \cite{GAprl}, the results of this analysis are interesting, both conceptually and from the point-of-view of  astrophysical applications. It is obviously interesting that the vortex-mediated mutual friction may lead to the presence of an instability. However, this is perhaps not surprising given the wealth of results relating to instabilities triggering superfluid turbulence \cite{turb1,turb2} and previous results in the context of pulsar precession \cite{fp1,fp2}, but the present analysis provides the first demonstration of this kind of instability for global mode oscillations. We are also breaking new ground by considering perturbations for background configurations with differential rotation.  It is notable that, as soon as we allow for this extra degree of freedom the problem becomes  richer.

\begin{figure}[h]
     \includegraphics[height=10cm,clip]{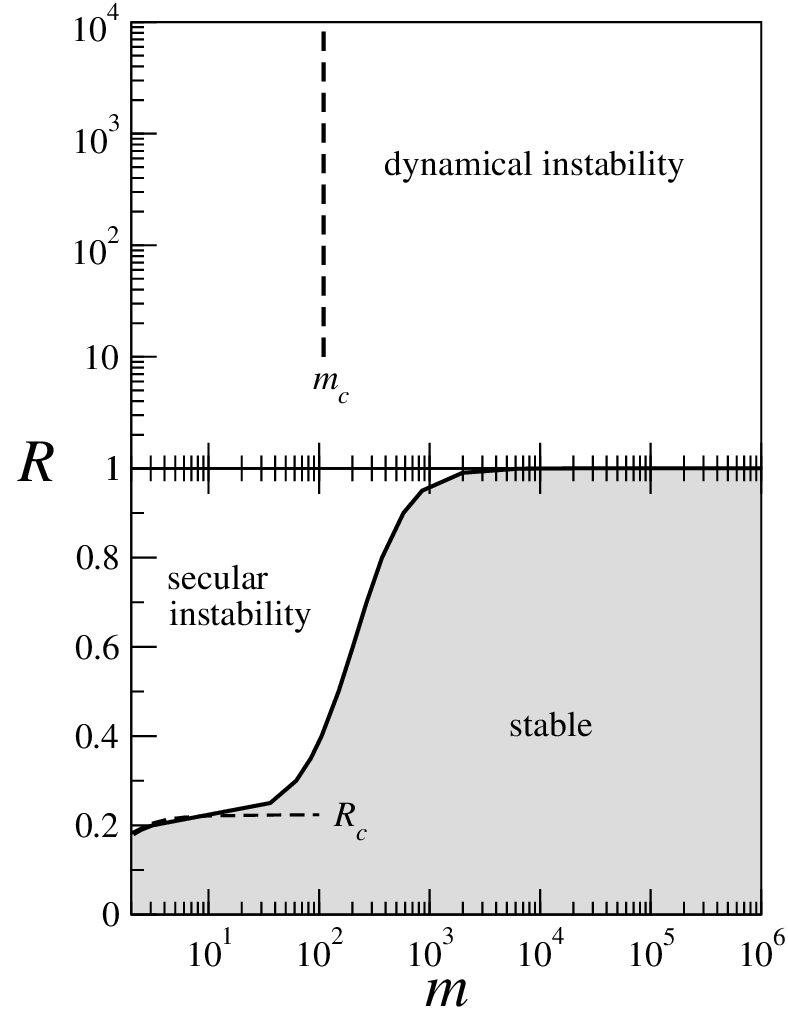}
     \caption{Summary of the parameter space  $\mathcal R$ vs $m$, indicating the different regions of instability for $x_\p=0.1$ and $\Delta = 5\times10^{-4}$. Dashed curves show the critical multipole $m_c$ where the behaviour changes from secular to dynamical instability on the strong coupling regime (upper panel), and the estimated critical drag $\mathcal R_c$ from \eqref{Rcrit} where the secular instability sets in for low $m$-multipoles (lower panel). }
  \label{pplane}
\end{figure}

A particularly interesting aspect of the results is the presence of both secular and dynamical instability behaviour in (more or less) distinct parts of parameter space. The behaviour that we have unveiled is summarized in 
 the phase plane  in Figure~\ref{pplane}, that shows the unstable region in the $m-\mathcal R$ plane. As far as we are aware, the present analysis represents the first detailed study of a problem that has secularly unstable modes entering a regime where they become dynamically unstable. The associated behaviour is, in many way, predictable. For example, instead of having modes becoming dynamically unstable as real-valued frequency pairs merge and form complex conjugate pairs, we now have dynamical instability behaviour associated with ``near misses'' in the complex frequency plane. Our results suggest that dynamical instability only results provided that the damping of the modes is not too large (in our case $\mathcal  R \ge 10$ or so). To improve our understanding further, we need to consider the necessary and sufficient criteria for the superfluid instability to operate. This is a challenging problem but it seems likely that one could make progress by adding the perturbed mutual friction force to  the results in \cite{canon}.

 An interesting analogous problem concerns the $l=m=2$ bar-mode in rotating neutron stars. This mode is known to become secularly unstable due to the emission of gravitational waves well before (along a sequence with increasing degree of differential rotation) it becomes dynamically unstable, see \cite{gwrev} for a review and  \cite{barmode} for recent work. The behaviour of that instability may be quite similar to that of the superfluid problem we have discussed here. If this is the case, then a key question concerns whether there always is a dramatic change in gravitational-wave emission rate before and after the critical parameter for onset of the dynamical instability is reached. In the bar-mode problem there is also another class of instability, commonly referred to as the low $T/W$ instability \cite{lowTW}. We have no evidence for an analogous instability in the present problem. 
   
Even though we have not discussed possible astrophysical repercussions in any detail, it is clear that 
the potential link with the unresolved problem of radio pulsar glitches provides strong motivation for further work on this problem. Of course, we cannot at this point really tell how relevant the new r-mode instability is in this context. In order to act in an astrophysical system, the instability must be robust enough to remain once we account for the star's magnetic field and the elastic crust.
These features are likely to affect the instability considerably, but more work is required if we want to quantify what the effects may be. The two-stream instability mechanism is sufficiently generic that it would be remarkable if it would cease to operate in more complex settings, but it could be that the instability threshold moves out of reach for a real system. At this point, we cannot say. There is, however, fresh evidence supporting 
 the existence of short wavelength superfluid instabilities in the presence 
 of a magnetic field and/or an elastic crust, see \cite{link1,link2}. 
 The exact relation between these instabilities and our r-mode 
 instability is still unclear, apart from the fact that both require
  vortex mutual friction to operate. The secular instability behaviour is obviously also worthy of further attention. In particular, since it is relevant also for large scale modes, including the lowest multipoles that are the most important from the gravitational-wave point-of-view. The problem is clearly extremely interesting and well worth returning to in the future.

\acknowledgements

NA is supported by STFC in the UK. KG is supported by the Ram\'{o}n y Cajal Programme in Spain. 


\end{document}